# To Aid Scientific Inference, Emphasize Unconditional Compatibility Descriptions of Statistics

Sander Greenland*
Department of Epidemiology and Department of Statistics,
University of California, Los Angeles, CA, U.S.A.
lesdomes@ucla.edu

Zad Rafi
555 S. Columbus Ave.
Mt. Vernon, NY 10550, U.S.A.

Robert Matthews
Department of Mathematics
Aston University, Birmingham, UK

Megan Higgs
Critical Inference LLC
Bozeman, MT 59715, U.S.A.

*Corresponding Author

Suggested Running Head: Emphasize Unconditional Compatibility Descriptions

Conflict of Interest Disclosure: The authors declare no conflicts of interest.

**Abstract:** All scientific interpretations of statistical outputs depend on background (auxiliary) assumptions that are rarely delineated or explicitly interrogated. These include not only the usual modeling assumptions, but also deeper assumptions about the data-generating mechanism that are implicit in conventional statistical interpretations yet are unrealistic in most health, medical and social research. We provide arguments and methods for reinterpreting statistics such as *P*-values and interval estimates in *unconditional* terms, which describe compatibility of observations with an entire set of underlying assumptions, rather than with a narrow target hypothesis *conditional* on the assumptions. Emphasizing unconditional interpretations helps avoid overconfident and misleading inferences in light of uncertainties about the assumptions used to arrive at the statistical results. These include not only mathematical assumptions, but also those about absence of systematic errors, protocol violations, and data corruption. Unconditional descriptions introduce assumption uncertainty directly into the primary statistical interpretations of results, rather than leaving it for the discussion of limitations after presentation of conditional interpretations. The unconditional approach does not entail different methods or calculations, only different interpretation of the usual results. We view use of unconditional description as a vital component of effective statistical training and presentation. By interpreting statistical outputs in unconditional terms, researchers can avoid making overconfident statements based on statistical outputs. Instead, reports should emphasize the compatibility of results with a range of plausible explanations, including assumption violations.

**Acknowledgements:** We are most grateful for generous comments from Nicole Lazar and Allen Schirm.

**Background**

We have elsewhere reviewed proposals to reform terminology and improve interpretations of conventional statistics by emphasizing logical and information concepts over probability concepts [1–3] . We here explore how common descriptions of statistical outputs can be technically accurate yet still mislead when there is considerable uncertainty about background assumptions (as usual in many contexts). Typical examples of background assumptions include: patient outcomes are independent; interactions and trends follow the regression model used for analysis (e.g., linear or logistic); there are no uncontrolled sources of systematic error (e.g., no uncontrolled confounding, subject-selection bias, measurement error, or sparse-data bias [4]); and there is no selective reporting of results based on their *P*-values, interval estimates, or any other output – or if there is such selection, any bias it could produce has been adjusted for (or blocked) by the computation and presentation of the final results.

Misleading interpretations arise because the descriptions are *conditional* on such assumptions (i.e., they treat them as *given*), and so do not factor uncertainty regarding the assumptions into their assessments. While various risk-assessment methods can incorporate additional sources in quantified uncertainties, such methods demand considerably more skilled user input than do conventional regression methods [5–9] and still do not account for unquantified sources of uncertainty. We thus present a more direct and nontechnical approach based on logic and associated language to honor assumption uncertainty, called *deconditioning*. Deconditioning shifts the logical status of assumptions in the description of results *from* what is assumed as given (conditioned on) *to* an integral part of what is being checked by common statistics. This shift makes clear that the results might be explained by violations of background assumptions rather than, or in addition to, failure of a targeted hypothesis. Such deconditioning does not change calculations, hence it does not change any *P*-value, interval estimate, or posterior distribution. Instead, it provides a different interpretation of these statistics, one arising naturally in thorough contextual discussions, but which deserves more emphasis in basic statistical practice, including teaching and presentations.

In philosophical accounts of theory testing, background assumptions correspond to auxiliary assumptions [10–13]. Our usage, however, explicitly subsumes assumptions of researcher competence and integrity in the conduct and reporting of statistical analyses (which have come under fire in the "replication crisis"), narrow and commonly stated statistical assumptions (such as regression linearity, independence, homogeneity in variance, etc.), and methodological assumptions (such as no uncontrolled confounding). Perhaps because of this generality, we have found that our recommendation to decondition inferences (present inferences using an unconditional interpretation) [1, 3] is often difficult for statistically sophisticated readers to comprehend, or at least buy into, and is even resisted, misunderstood, and misrepresented by some reviewers with extensive credentials in statistics and philosophy.

The resistance likely reflects how deeply engrained making inferences conditional on all the aforementioned types of assumptions is to current norms of statistical practice. Thus, the present paper explains at length our practical rationale for de-emphasizing conditional interpretation, such as those based on terms and concepts such as "significance" and "confidence", in favor of unconditional interpretations framed in terms of compatibility between data and proposed models of data generation, as seen discussions of *P*-values for model checking [14–16]. In the context of de-conditioning, the "models of data generation" include all background assumptions. The present paper is thus a sequel to our previous recommendations for improving scientific interpretations of statistics [1] and is meant to convey the core ideas to users of statistics in scientific research, as well as provide shared working language and notation for the concept, rather to satisfy technical or philosophical demands; for the latter, Greenland [17] provides a complementary theoretical treatment. This paper also extends in detail some of the reasons behind calls to abandon the use of terms like "significant" when referring to *P*-values and in tandem cease dichotomization of *P*-values into "significant" and "nonsignificant" categories [18–22].

**Deconditioning to Prevent Uncertainty Laundering**

Treating formal statistics as if they capture all important uncertainty sources has been labeled *uncertainty laundering* [23], which is exactly what is done whenever discussions revolve around whether results are "statistically significant" or whether interval estimates contain a null value. Even when such laundering is not performed, expressions of uncertainty about most background assumptions are usually reserved for informal discussion, which often takes the form of "walking back" conditional descriptions provided in an earlier results section [24].

We advocate instead the deconditioning of inferences by shifting to unconditional descriptions. This deconditioning introduces assumption uncertainty directly into the primary statistical descriptions of results, to acknowledge that the results cannot claim to have captured all uncertainty if there are any reasonable doubts about the assumptions or models used to derive the statistics. We thus view use of unconditional descriptions as a vital component of effective statistical training and honest statistical presentation. It is a deviation from conventional practice that deserves more of the spotlight in the movement to reduce misinterpretations found in teaching materials and research articles and reinforced through teaching. It also provides a bridge from traditional statistics to modern approaches based on causal logic and bias analysis.

**The central role of causal explanations in interpreting statistics**

To understand common statistical summaries in their scientific context, we must imagine an open-ended list of contextually plausible mechanistic (causal) explanations for an observed statistic falling where it did. With conditional interpretations, the only explanations allowed from that list are those adhering to the background assumptions used to compute the statistic. This means that conditional explanations assume a very restricted list, one which allows only combinations of random error and violation of a statistical target hypothesis **H**, rather than violation of background assumptions. This target **H** is an assumption that has been singled out for testing; typically, **H** states that a single parameter of a model takes on a "null" or test value, although in model "goodness-of-fit" tests it refers to a vector of parameters. The statistic and its conditional interpretations are mute about reasonableness of the background assumptions in the application context.

Because conditional interpretations ignore uncertainty about background assumptions, they aggravate fallacies such as treating the "nonsignificance"/"significance" dichotomy as a true/false indicator for the scientific hypothesis to which **H** is supposed to correspond rather than the decision indicator it is. A common example arises when the scientific hypothesis is a causal null hypothesis that some treatment or exposure has no effect on a particular outcome. **H** is then a statistical null hypothesis that the treatment has no association with the outcome within an assumed background model that, among other things, purports to adjust for (block or eliminate) bias sources, and the "significance" declaration is only a decision to proceed as if the treatment has an effect [25]. Among possible sources of bias are causal pathways that contribute to the association of treatment with the outcome beyond the association induced by the treatment effect on outcome [26].

When model-based adjustment fails to block all bias sources (as is likely when background assumptions are violated), violation of the scientific hypothesis no longer corresponds perfectly to violation of the statistical hypothesis **H** [27–29]. Consequently, because claimed properties of statistical procedures (e.g., rejection and coverage rates) pertain only to the statistical hypothesis **H**, such properties no longer apply to the hypothesis of actual scientific interest. This is often called the problem of residual bias due to misspecification of the sampling model. For example, when age effects are modeled using only broad age categories, the model includes an implicit background assumption that age has a negligible effect within the categories; if this is not so (as with cancer and Covid-19 risk) the association tested in **H** may correspond poorly to the causal effect of scientific interest (often called a problem of residual age-confounding), and the actual error rates of the scientific decisions will be far from their nominal values (since those rates apply only to decisions about **H** when the model is correct) [30, 31].

In contrast, an unconditional interpretation does not assume the adjustments based on the model eliminated all bias sources, nor does it equate violation of the statistical target **H** with the violation of the causal target. Instead, unconditional inference starts with a list of plausible causal explanations for observed associations, including nonrandom physical (causal) mechanisms that violate background assumptions other than **H** (e.g., mechanisms that produce nonrandom selection and nonrandomly missing information [29]). In typical social-science and biomedical applications, there will be multiple such explanations. These explanations will not be mutually exclusive, and some may not even be within the awareness of researchers (e.g., undetected programming errors).

Accurate discussion sections of articles do consider a list of such explanations along with possible background evidence bearing on them, although this list typically comes *after* conditional inferences were provided earlier in the article. Our point is that the spectrum of possible explanations for observations maps onto possible violations of auxiliary (background) assumptions (beyond mere violation of **H**). Thus, unless every such assumption is beyond the reasonable doubt of all stakeholders (which is rarely the case), unconditional interpretations should be provided as the primary interpretation of the statistical, not as an afterthought to “walk back” a previous conditional interpretation. Again, this involves a shift in expectations, but no change to how statistical summaries are calculated. This approach explicitly acknowledges the existence of uncertain background assumptions, but it does not require delineating all assumptions (which is typically impossible) or evaluating the extent to which each is violated.

**An Example**

As in our previous paper [1], we will illustrate problems and recommendations with published results from a record-based cohort study of prescriptions for serotonergic antidepressants (selective serotonin reuptake inhibitors, SSRIs) during pregnancy and subsequent autism spectrum disorder (ASD) of the child [32]. That paper first reported an adjusted ratio of ASD rates (hazard ratio or HR) of 1.59 when comparing mothers with and without the prescriptions, and 95% "confidence" limits (CI) of 1.17 and 2.17. This estimate was derived from a proportional-hazards (PH) model which included maternal age, parity, calendar year of delivery, neighborhood income quintile, resource use, psychotic disorder, mood disorder, anxiety disorder, alcohol or substance use disorder, use of other serotonergic medications, psychiatric hospitalization during pregnancy, and psychiatric emergency department visit during pregnancy as predictors of the child's outcome (ASD).

The paper then presented an analysis with adjustment based on a high-dimensional propensity score (HDPS), which included the above variables and more as predictors of the treatment (maternal SSRI use in pregnancy). The estimated hazard ratio became 1.61 with a 95% CI spanning 0.997 to 2.59. The point estimate corresponds to a 61% increase in the hazard rate in the exposed children, and the interval estimate includes ratios as large as 2.59 and no lower than 0.997. The lower limit rounds to 1.00; yet, because 1 is just inside the interval, the paper declared that "*in utero serotonergic antidepressant exposure compared with no exposure was not associated with autism spectrum disorder in the child*." Although the authors also said a causal relationship cannot be ruled out, their main conclusion of "no association" was a misinterpretation of their own results insofar as the observed association was much closer to the 70% increase they reported from other studies [33] than to no association (HR = 1) [1].

An alternative and more appropriate (albeit still conditional) description of the results would be

> After HDPS adjustment for confounding, an estimated 61% hazard elevation remained; however, under the same model, every hypothesis from no elevation up to a 160% hazard increase has a P-

> value above 0.05. Thus, while quite imprecise, these results are consistent with previous observations of a positive association between serotonergic antidepressant prescriptions and subsequent ASD. Because the association may be partially or wholly due to uncontrolled biases, further evidence will be needed for evaluating what, if any, proportion of it can be attributed to causal effects of prenatal serotonergic antidepressant use on ASD incidence.

This alternative description provides an example of conditional interpretation, followed by a statement of the limitations of the conditional statement provided immediately before. To move closer to an unconditional interpretation, at a minimum the following warnings should be added to the description:

> Any clinical advice regarding the continued use of serotonergic antidepressants in pregnancy will need acknowledgment and quantification of the uncertainties not captured by the statistical results, as well as of the potential costs and benefits of switching to available alternative medications.

We believe such caveats should always accompany summaries of statistical analyses, to highlight their limitations as a basis for decision-making.

**Summary Statistics and their Reference Distributions**

Central to the present treatment is the idea of "standardized" distances, which are measures of discrepancy or divergence between a summary of the observed data set and the data expected when all model assumptions hold, scaled to account for hypothesized random fluctuations. This idea is seen for example in Pearson's $\chi^2$ test [34], in which the squared differences between observed and model-expected counts are scaled to their expectations to create a squared Pythagorean distance between the vector of counts and the vector of expectations in an expectation-scaled space. Traditionally, such a summary divergence measure is called a "test statistic". Rather than identify this statistic with a statistical test, we treat it as a continuous measure of geometric divergence of the data from the model, whose units are derived from its reference distribution (such as a normal, t, or chi-squared distribution).

That distribution is usually interpreted as the limiting relative histogram of the statistic that would be seen upon hypothetical repeated sampling of the data when all the assumptions about the sampling process are satisfied (including the tested assumption **H** as well as the background assumptions). An alternative to this conventional repeated-sampling view instead treats the entire collection of data sets that the assumed sampling process could generate (the sample space under the model when the data are fine enough to represent equiprobable points) as a single fixed entity [35]. The reference distribution for the statistic is then the distribution of the summary statistic over this sample space, which includes data sets that *could have* been generated from the process (counterfactual data) as well as the observed data set. This subtle shift focuses on where the observed statistic lies within the reference distribution derived from the model assumptions, without depending on notions of infinite exchangeable repetitions that vex practical-minded students (who will rarely see more than one such repetition of a given study).

**Reference Distributions versus Realities**

Unfortunately, in current practice in many disciplines, the reference distribution is simply a default in the statistical package used to process data into desirable output for publishing, such as *P*-values, "significance" declarations, and "confidence" intervals. The derivation of these statistics is usually detached from the actual research setting (including deviations of the study generating the data from the assumed hypothetical randomized experiment underlying the default reference distribution). Reference distributions with little or no grounding in typical research realities are now routine, and that is one practical justification for the unconditional approach.

Our primary concern is that when the background assumptions used to derive reference distributions are violated, actual error rates of tests or intervals may be far higher than the stated rates; in that case, declarations of "significance" or "nonsignificance" are deceptions, and "confidence" intervals become *overconfidence* intervals [3, 36]. Specifically, in field studies and studies of human subjects, the statistical model is hypothetical, for it is never the case that *all* the background assumptions it contains hold; there are always study problems, and it is implausible that any statistical model we use (whether for outcomes or exposures) is correct or complete in all respects. Even if we list potential problems, there will inevitably be additional violations we are unaware or dismissive of, or discouraged from listing, such as data corrupted by file-handling errors or even outright fraud [37–39]. But such possibilities should not be overlooked, for even modest violations of background assumptions can easily bias a *P*-value toward 0 or 1, thus pushing them over cutoff (alpha levels) – regardless of whether the target hypothesis is correct – and thus invalidating error-rate claims and decisions based on the conditional interpretation of that *P*-value.

As elsewhere [1-3], we will consider only statistics that satisfy the most basic frequentist validity criteria. In particular, a *P*-value is uniformly calibrated (or valid for statistical testing) if all possible values for it from zero to one are equally likely (uniform in probability) when both the statistical hypothesis **H** targeted by the statistical test *and* all the background (auxiliary) assumptions **A** used to compute the results hold. With this validity criterion met, we can also correctly describe the *P*-value (expressed as a percent) without explicit reference to repeated sampling or randomization, as the *percentile* at which the observed divergence statistic falls in the distribution for the statistic under the target hypothesis *and* the background assumptions [35, 40].

The purpose of the percentile description is to connect the *P*-value to a familiar concept, the percentile at which someone's score fell on a standard test relative to all other test takers (e.g., a college or graduate admissions examination), instead of the more remote abstraction of repeated sampling from a hypothetical population or model. The general idea of "percentile in a reference distribution" helps make a concrete connection to everyday experience, whereas infinite repetitions are infinitely far removed. It also aids in incorporating directly into statistical logic all the possible explanations of why a study's observed statistic fell where it did and thus why a *P*-value ended up where it did. Just as we would consider why someone's test score fell where it did, including amount of preparation and sleep as well as "bad luck", we should consider why a statistic fell where it did, including nonrandomness in allocation, adherence, and measurements, as well as "random error". As will be reviewed below, the percentile interpretation also provides a natural extension to interval estimation in the form of compatibility intervals [1-3, 17] .

**Some Technical Asides on Statistical vs Logical Usage**

We emphasize that our use of the terms "conditional" and "unconditional" is in the logical and common sense of assuming vs. not assuming certain conditions hold when making statements about the application; for example, contrast the conditional "If this study is valid then the drug must be effective" against the unconditional "Either this study is invalid or else the drug must be effective". The conditional statement restricts interpretation *given* the assumptions hold, while the unconditional statement considers the possibility of the assumptions not holding as part of the interpretation. Our logical use of the terms should be distinguished from usage in mathematical statistics, such as the distinction between probability calculations that restrict the sample space based on observed events vs those that do not (e.g., as in conditional vs. unconditional tests in 2 x 2 tables), or between conditional vs. unconditional model parameters (e.g., as in subject-specific vs. group-level parameters in models for clustered data).

Conventional treatments consider **H** to be composite when there are explicit nuisance parameters in the model; a search is then made for a reference distribution that follows from all models in the class generated by varying the nuisance parameters. Complications arise when, as often in the composite case and in small samples, the resulting *P*-values turn out to be "conservative" (stochastically larger than a uniform random variable) [17]. When, however, we further allow that the model may be wrong, we may recognize that a *P*-value refers to the entire class of models that generate or lead to the reference distribution from which it is computed, and thus in the logical sense **H** is always composite: It includes many possibilities that do not follow the original scientific hypothesis but mimic its predictions for the study under analysis [17, 27].

**Deconditioning by explication of alternative causes of observations**

As is well known, the presence of an association in multiple observational studies does not by itself mean that the association is causal (e.g., that the drugs under study cause autism). In fact, Brown et al. [32] argued that the associations seen in their initial results represented confounding

– a spurious association due to an association of SSRIs with the actual causes of ASD. But such a confounding hypothesis should not be confused with lack of association; instead, it should be treated as one of several possible explanations (ranging from real drug effects to random error), any or all of which may be contributing to the observed association.

The statistical adjustments used by Brown et al. [32] were in fact designed to minimize confounding, and thus they greatly diminish its plausibility as a major source of the observed association. Furthermore, some of the suggested explanations might have *reduced* the observed association; in particular, in large enough samples, random error is just as likely to deflate as to inflate an observed association. Without study design features to block alternative explanations (e.g., randomization to minimize confounding), statistical results cannot settle these matters. We thus need statistical descriptions that do not read as if they are decisive, but instead contain language that acknowledges model uncertainties and possible alternative explanations.

Even when technically correct, common interpretations of statistics are deficient in this regard. Consider that a $P$-value $p = 0.0625$ yields a binary $S$-value (surprisal) $s = -\log_2(p) = 4$ bits of information against the target hypothesis **H**, *if* all the background assumptions are correct [1, 3]; see also the Appendix. This description is *conditional*, in that it assumes an entire set **A** of background conditions in order to compute and interpret $p$ and $s$. Taken together with **H**, these background assumptions **A** compose the underlying model used to compute the $P$-value $p$ and $S$-value $s$. Without changing any computation, we can reinterpret $p$ and $s$ unconditionally, with $p$ showing the compatibility of the data with the analysis model and $s$ showing the information this $p$ provides against this model. The following sections elaborate this view in more logical detail.

**Hypotheses, models, and their statistics**

A target (or test) hypothesis **H** may assert that an effect measure is a specific value (usually but not necessarily zero), or falls in a specific direction (as in noninferiority hypotheses), or falls in or out of a given interval (as in minimal-effects and equivalence tests); or it may assert that a given model falls within a specific model family (as in tests of fit). We will refer to the combination of the target hypothesis **H** *and* the collection of underlying background assumptions **A** used to derive statistics about **H** as the *target model* **M**. The collection of assumptions represented by **A** is used not only by traditional significance tests but also by their Bayesian analogs [16, 41]. With this notation, we can graphically display the distinction between conditional and unconditional interpretations as in Fig. 1, which shows how the conditional interpretation (**a**) targets only the hypothesis **H** under the dubious condition that there are no violations of background assumptions, whereas the unconditional interpretation (**b**) targets the entire set of assumptions used to compute the statistical results.

Figure 1. **Conditional versus unconditional interpretations of *P*-values, *S*-values, and compatibility intervals (CIs)**: (**a**) Conditional interpretation, in which background model assumptions, such as no systematic error, are assumed to be correct; thus, the information provided by the *P*-value and *S*-value is targeted towards the hypothesis **H**. (**b)** Unconditional interpretation, in which no aspect of the statistical model is assumed to be correct; thus, the information provided by the *P*-value and *S*-value is targeted toward the entire model **M**, which includes all auxiliary assumptions in **A**.

In mechanistic terms, possible nonrandom causes of an extreme divergence between the data and the model (and thus a small *P*-value) include not only the target hypothesis **H** being wrong, but also or instead some other violation of one or more assumptions contained in **A**. Conversely, possible nonrandom causes of a large *P*-value include not only the targeted hypothesis **H** being correct, but also assumption violations that mask a true deviation from **H**. These problems remain and may even be intensified by choosing a test statistic whose sensitivity (power) is maximized for violations of **H** conditional on the list of auxiliary assumptions **A**. Because such conditional interpretations focus attention away from **A**, small *P*-values from such "optimized" or "most powerful" statistics are easily misattributed to violation of **H** while large *P*-values are easily misattributed to correctness of **H**, rather than to possible violations of auxiliary assumptions in **A**.

We thus can and should view the *P*-value as referring to a probability derived from the entire target model **M** composed of **H** and **A**, and the *S*-value as measuring the information supplied by the divergence statistic against **M**, even when the statistic is optimized for violations of **H** given **A**. This description is *unconditional* because it places the background assumptions in the spotlight with the target hypothesis **H** when considering explanations for results, rather than leaving them in the background. Unconditional interpretations explicitly state that violation of auxiliary assumptions rather than **H** may be responsible for the results, with no conditions imposed. For example, a typical yet commonly overlooked assumption in **A** is that there was no selection of models or methods based on looking at the statistical outcomes. Under that assumption, selection of models that yield narrow intervals for "higher precision" will cause an excess of small *P*-values even if the target hypothesis **H** is correct. Then too, selection of models or methods on the basis of yielding wider intervals for "conservative inferences" (a form of null bias) can cause an excess of large *P*-values even if the target hypothesis is false.

More generally, the smaller the *P*-value and thus the larger the *S*-value we observe, the more justification we have for saying that it appears one or more assumptions in the target model are wrong. This unconditional analysis does not however indicate *which* assumptions are wrong. The reasons for the assumption violations might include the target hypothesis **H** being false, but may also include uncontrolled bias, data-collection errors, programming errors, data tampering or fabrication [37–39], or some other deviation from the background assumptions hidden in traditional interpretations. This information limit of statistical analyses is inherent and universal, but underappreciated in practice; a notable example is the report of faster-than-light neutrinos which turned out to be due to equipment defects [42].

In parallel, if we observe a large *P*-value and thus a small *S*-value, we cannot conclude that there is no violation of any assumption; quite contrarily, it may be that the assumption violations biased the *P*-value upward instead of downward. This caution is just the unconditional version of the warning dating back to Pearson in 1906 [43] and often repeated since [36, 44–46]: A large *P*-value is **not** evidence that the target hypothesis is correct. As reflected by the small *S*-value, it simply means the divergence statistic supplied little information against the target hypothesis *or any other assumption used to compute the P-value*. This lack of information reflects the limitations of the statistic (which in turn may reflect limitations of the study), and thus should not be taken as support for the absence of an effect [45].

**The importance of unconditional interpretations for scientific practice**

The unconditional interpretation is particularly important in contexts such as health and medical sciences, where researchers can rarely, if ever, achieve full control over all potential sources of systematic error. Even randomized trials can suffer from systematic errors due to drop-out, censoring, protocol violations, and other problems [47, 48], and so are not the "gold standard" they are sometimes claimed to be. In contrast, typical physical-science experiments may control all important conditions and so justify a conditional interpretation (although again, serious exceptions occur even in particle physics [42]). We thus suspect that the nearly exclusive emphasis on conditional interpretations seen in health, medical and psychosocial sciences is an inappropriate emulation of practices in disciplines that can achieve far more experimental control than possible with human subjects (e.g., physical sciences). This emulation reflects the ritualistic and mechanical use of statistical methods in current scientific culture [49].

We view explication of the conditional vs. unconditional distinction as crucial to effective teaching, and the unconditional view as essential for principled and honest practice: When (as usual in our experience) there is meaningful doubt about the assumptions underlying a statistical procedure, we need to remind ourselves of the unconditional fact that any result ("large" or "small") may have occurred not only from "chance" but also from assumption violations. Such reminders are seen in well-reported studies, which caution about possible sources of bias in the study, although usually as a set of disclaimers following the conditional interpretation of results, rather than an integral part of the primary reporting of those results. We thus hold that unconditional interpretations should be presented when (as usual) any reasonable doubt can be raised about background assumptions, and should become the primary interpretations covered in statistical education.

The unconditional interpretation is far more helpful than the conditional when there are concerns about violations of assumptions, including protocol violations. Suppose for example there are plausible concerns about violations of the data collection, processing, or reporting protocols. A common concern is that a result was selected for special emphasis out of several, or many, based on its associated *P*-value (whether for being high, "downward hacking"; or low, "upward hacking"), or one CI of several was emphasized while others were downplayed based on including vs. excluding a null value. For example, contrast the discussion of the results from HDPS versus the proportional-hazards model in Brown et al. [32]. Worse, some results go unreported based on their *P*-values or CIs, thus becoming nonrandomly missing information. A conditional interpretation assumes there is no such uncontrolled selection of summaries based on what they favor, and so is misleading when selection is a possibility. In contrast, an unconditional interpretation will list uncontrolled selection bias among the possible causes contributing to (that is, partial explanations for) the observed *P*-value or CI.

The multiple explanations allowed by the unconditional view show why in that view it would be fallacious to simply say that a small *P*-value favors an alternative hypothesis or an *S*-value measures the information supporting that alternative. Considering the example, it would be wrong to say the *S*-value of 4.31 against the no-effect hypothesis (that the drug does not affect risk) measures the information favoring the causal alternative that taking the drug increases risk: Such an interpretation would have to assume that the 61% higher rate seen with the drug is solely a product of genuine drug effects and random errors, which is not credible due to the possibility of systematic errors from violations of background assumptions (such as the assumption of no uncontrolled bias). In general, shifting to use of unconditional interpretations is a way to make scientific limitations of common statistical summaries explicit, and to prevent lending undeserved confidence or authority to the summaries (as seems to routinely afflict conditional interpretations).

**Conditional versus unconditional compatibility**

To summarize thus far: an unconditional interpretation does not take the background model as correct, and thus cannot support the usual rationales for *P*-values and CIs based on error rates for test decisions or coverage. In contrast, compatibility interpretations can be viewed in several logically conditional ways, or in an unconditional manner.

To illustrate, consider the usual Cox proportional-hazards (PH) model in which the target is a treatment coefficient $\beta$, where $e^{\beta}$ is the hazard ratio comparing event rates in treated versus the untreated. Among the background assumptions **A** of this model are that hazard ratios remain constant across time and covariate levels. Suppose **H** is the additional model constraint $\beta = b_H$. A *P*-value $p_H$ for **H** may then be derived by combining **A** and **H** to produce a model **M** in which $\beta$ is $b_H$ exactly, then using the data to fit **M** and compare it to the fit of the initial model defined by **A**.

As a compatibility measurement, $p_H$ is usually described conditionally as indicating the compatibility of **H** with the data given the background assumptions **A**. Nonetheless, $p_H$ can also be described instead as indicating the compatibility of **H** with **A** given the data, when compatibility is measured by how much the addition of **H** changes the expected (fitted) data from the expectations obtained using **A** alone. This data-conditional interpretation has direct connections to Bayesian results [16]. It can also be described as indicating the compatibility of **H** with the combination of **A** and the data, which is an unconditional interpretation.

The differences among these statements may seem subtle, but are profound. Conditioning on **A** can enable error-rate interpretations (such as claims of "statistical significance" and "coverage") that appear much stronger than mere compatibility assessments. But when **A** is uncertain, those conventional interpretations assume what is **not** known to be the case, rendering them at best an interpretation for a hypothetical world in which **A** holds – and which may be far from reality. This hypothetical quality gets overlooked when, as usual, the background model **A** is left out of descriptions of the statistical results, thus making those results seem more definitive than they are. Compatibility interpretations can however be limited to what has been observed, treating the statistics as descriptions of relations among hypotheses, models, and data [17]. Furthermore, they allow conditioning on the data alone, which is less misleading insofar as it is no longer conditioning on something whose status is largely uncertain – albeit not completely certain, since the data themselves may be corrupted [42]. The unconditional statement has the advantage of taking nothing for granted; even bare procedural assumptions (e.g., that the data really are as given and computations were done correctly) can, and should, be listed and placed in **A**.

**Compatibility and Deconditioning as Cautious Description**

As with the logically much stronger error-rate interpretations, compatibility interpretations are framed around a targeted hypothesis **H** and the background **A**. The statistic used is typically chosen to measure and thus be sensitive to data deviations from the target **H** given the background assumptions **A**; in this sense the *derivation* of the statistic is logically conditioned on **A**. An example is the common Wald or Z-statistic for evaluating $\beta = b_H$, $|b_A - b_H|/sd$, where $b_A$ is the estimate of $\beta$ obtained when only **A** is assumed and sd is its estimated standard deviation. As with traditional interpretations of statistics (e.g., “statistical significance”, “confidence”, or decision rules) this sensitivity optimization for **H** also results in sensitivity to violations of **A** that mimic or mask violations of **H**; thus, interpretation of results should always be cautioned as dependent on the background model **A** and the statistic chosen to measure divergence of **H** from **A.**

One may object that, even unconditionally, compatibility interpretations will still be biased by assumption violations, where the bias may be toward understating or overstating compatibility. But that objection reflects a deep misunderstanding of the meaning of compatibility: “High compatibility of **H** with **A**” merely says the summary statistic chosen to measure compatibility places the model **M** (which combines **H** and **A**) is near the model **A**; *it makes no claim whatsoever that violations of* **A** *are absent or that* **H** *is acceptable scientifically*. Conversely, “very low compatibility” (or “highly incompatible”) merely says the statistic places **M** and **A** far apart; *it makes no claim whatsoever about whether violations of* **A** *or* **H** *or both are responsible for the divergence of* **M** *from* **A**. Compatibility allows us to remove conditioning on **A** and thus make statements that do not depend on shaky assumptions. This deconditioning is a constructive response to the maxim “absence of evidence is not evidence of absence” [45], as it involves no inference about *why* **H** and **A** appear compatible or incompatible; such inference must be left to evaluation of the assumptions in **A** in terms of possible causes of violations [27].

In general, compatibility interpretations refuse demands for conclusive assessments, even of uncertainty. The core idea is that the conventional statistics demanded for publication (mostly *P*-values and frequentist estimates) can only gauge incompatibilities between our data and the models we use to analyze the data, or between different models in light of the data. Those statistics may falsify certain models by revealing gross incompatibilities between them and the data. The converse is incorrect, however, because compatibility alone should never be treated as support for the model. And because there is no restriction on how model violations may occur, the low compatibility of one model with the data does *not* provide support for a competitor. Each alternative model needs to be evaluated directly against the data, with its own *P*-value and *S*-value, as well as being evaluated relative to one another (as in pure likelihood and Bayesian statistics).

In summary, compatibility interpretations are far more reserved and cautious compared to conventional error-rate descriptions. In a parallel fashion unconditional descriptions are far more reserved and cautious compared to conditional descriptions. While this caution may seem excessive, the usual conditional error-rate interpretations are nothing more than thought experiments unless there is empirical evidence *against* mechanisms that lead to assumption violations. And in one respect, deconditioning is not cautious enough, because it is no substitute for model diagnostics such as residual plots and direct measures of model fit to gauge the nature and extent of discrepancies between models and data.

Regardless of type of statements, ideal practice should be listing and critical evaluation of background assumptions, with clear rationales for each one. Unfortunately, some assumptions (such as no unreported model selection [50, 51]) cannot be checked by the reader, while other assumptions cannot be checked (i.e., are nonidentifiable) even if we are given the study dataset and full details on how it was collected. For example, the hypothesis that an observed association (or lack thereof) was due to confounding by an unmeasured variable cannot be assessed without assumptions about the relation of that variable to observed variables (assumptions which might be based on data external to the study under analysis). Thus, when important uncontrolled confounding is considered a serious possibility, an unconditional interpretation will refer to the observed association as an adjusted association rather than an "effect estimate", because the latter term invites conditioning on the assumption that the analysis successfully adjusted for all important confounding. In doing so, an unconditional interpretation will *not* deny the presence of an association simply because it is not "statistically significant" according to some model and cutoff.

**Compatibility intervals do not make coverage claims**

A "confidence interval" is often defined as an interval that contains the true parameter value in a percentage (usually 95%) of hypothetical study repetitions over a "long run" in which data variation follows some specified family of probability distribution. A more accurate term for such an interval would be "conditional coverage interval", since its coverage rate (and hence any confidence that supposedly follows from that) depends entirely on this assumption about data variation, which is part of **A**. Most statistical texts and critiques write as if these interval estimates are *only* defined or justified by their long-run coverage properties given the background assumptions [52]. If that were so, there would be no justification for the intervals outside of conditions that strictly enforce the coverage properties – conditions that are called into question whenever there is uncertainty about the background assumptions in **A**.

Fortunately, an interval labeled a “confidence interval” can be reinterpreted without a coverage requirement using unconditional interpretations, provided we reinterpret it as a *compatibility interval*, which retains the abbreviation “CI”. Specifically, we can bypass the need for a coverage interpretation by using the complementary mathematical relation between *P*-values and CIs. A CI of a particular level, say 95%, for a parameter $\beta$ summarizes the results of varying the target hypothesis **H**: $\beta = b_H$ over a range of possible values $b_H$ for $\beta$, and computing the *P*-value for each; the CI will then contain all values $b_H$ with a *P*-value $p$ exceeding $\alpha = 0.05$ [1-3, 53]. Thus, the CI contains a range of possible values $b_H$ that, in light of the data, are more compatible with the background assumptions **A** than are values outside the interval, and which would make the data less surprising given **H**: $\beta = b_H$ and **A** than would values outside the interval [3, 41]. This is so, regardless of the particular cutoff $\alpha$ chosen for the interval, *and regardless of long-run coverage* of the true value of $\beta$ by such intervals. We thus refer to a CI as a *compatibility* interval rather than a “confidence” or coverage interval [1-3, 17, 36, 53, 54], as this shift allows us to interpret the interval without assuming that **A** is correct, and thus decondition our description of the statistical results.

For exceptional situations in which **A** has been forced to hold by precise experimental design and conduct, a 95% CI can also be interpreted conditionally as in conventional interpretations, in that it shows the values $b_H$ of $\beta$ that would be unsurprising given **A** and the data in the sense that their *S*-value (surprisal) $s$ less than $-\log_2(0.05) = 4.32$ bits [3, 36].

The use of intervals introduces the issue of arbitrariness of the cutoff α for inclusion. That α is almost always the 0.05 (5%) default found in statistical software or to meet journal guidelines, corresponding to the default 95% "confidence level". Even in rare cases in which such intervals can be assured to have 95% coverage or a 5% error (noncoverage) rate (as in simulation studies [55]), there is almost never any contextual justification for using the 5% default. To accommodate other preferences, one could present CIs at varying levels (50%, 75%, 90%, 95%, 99%) and inspect the parameter values that fall within the bounds of those CIs [56]. Another option is to present a table of *P*-values for various possible effect sizes, including but not limited to the null [1]. If the number of possibilities is large, a graph of the *P*-values against the possibilities will produce a compatibility curve or *P*-value function (also known as a confidence curve, confidence distribution, or consonance curve) [1, 56–62].

In summary, just as Neyman-Pearson tests and confidence intervals are inversions of one another (a trivial corollary of their definitions), compatibility intervals are inversions of compatibility *P*-values. But unlike Neyman-Pearson statistics, compatibility is not forced into what are often useless or misleading conditional statements about error rates of decision rules or coverage. Compatibility intervals or regions are only shorthand summaries of compatibility measurements on models in a family. Those models share all their background assumptions **A**; the interval is constructed by measuring compatibility along one or a few dimensions of the model subspace defined by the family (e.g., the one defined by different values of a model coefficient). The intervals convey which models in the defined family meet some conventional descriptive minimum compatibility (usually $p > 0.05$) along that dimension.

**Compatibility intervals do not make confidence, credibility, or uncertainty claims**

As with "confidence" intervals, Bayesian posterior intervals (or "credible intervals") can also be reinterpreted as compatibility intervals, showing the parameter values most compatible with the background assumptions **A** given the data, where **A** now includes the prior distribution [1, 3]. As noted by some [65], if the data, model, prior distribution, or any other assumption in **A** do not inspire confidence or credibility, the interval should not either.

It follows that we regard recent attempts to relabel both "confidence" and "credibility" intervals as "uncertainty" intervals to be misleading, because these intervals usually do *not* capture major uncertainties about the data-generating process, and the traditional conditional interpretations do not convey this inherent limitation [17, 27, 53]. Capturing all important uncertainty sources requires more extensive subject-matter knowledge and modeling of the data generation process than most investigations have the time and expertise to conduct [5-8]. We thus conclude that, as with "random error only", the label "uncertainty interval" is a potentially misleading substitute for deconditioning. In contrast, a compatibility interpretation does not depend on how correct or incorrect the model assumptions are, nor does it claim to offer a sufficient accounting of sources of uncertainty; it is just a mathematical statement about a relation between the data and the model, however questionable the data or model may be.

**Compatibility intervals do not make power or accuracy claims**

An attempt at expressing caution about background assumptions is to describe CIs as measuring only the possible random error in the results; this attempt is inadequate because the CIs are computed using those assumptions [59]. Nonetheless, when CIs are very wide, they may suffice in conveying the inadequacy of the study for forming useful conclusions (beyond perhaps "more research is needed"). For example, a randomized trial producing a CI for a hazard ratio ranging from 0.90 (a 10% rate decrease) to as high as 20 (a 20-fold rate increase) should be taken to indicate that the results are too noisy to pin down even the direction of the association, even when **A** is given. This view leads to use of interval estimates to plan studies based on desired interval width [63, 64] rather than on statistical power (in which one fixates on the rate of rejecting the target hypothesis **H** under some specific alternative to **H** and the same questionable background assumptions). The goal is then to ensure that the region of compatibility above a given level is narrow enough to make the study reasonably informative *under the background assumptions* **A**. As with power analysis, that goal does not address violation of **A**; because those violations can drastically reduce the informativeness of the study and inflate error rates far beyond stated (nominal) levels, any claims that a proposed study will be accurate must be supported by design protocols that force **A** to hold [6].

In summary, in the unconditional compatibility view, any Neyman-Pearson, likelihoodist, fiducial, Bayesian or other claim about statistical power of a test or accuracy of an interval involves an act of faith in the model family (assumption set **A**) used to derive the test or region. That family will include the models used for random-effect or prior distributions, as well as the regression models used for treatment probabilities and causal models for outcome probabilities. If the resulting CI were empty or incredibly narrow, that would tell us that no model in the family met the minimum compatibility criterion. The only practical inference we could justify from such a result is that the model family defined by **A** has shown itself unsuitable for the scientific problem. When we are uncertain about **A**, a very narrow interval warns us that errors from model misspecification are no longer negligible compared to the random errors allowed by the model, and we should move on to more general model families that better account for real contextual uncertainties. This is easier said than done, the pressure to get precision can decrease motivation to consider models with more realistic background assumptions. Note that model diagnostics can fail us in this task because they too are compatibility measures, and like all statistics are incapable of discriminating among the infinitude of model families that appear compatible with the data (as may be seen from the many ways one can “overfit" data).

**Discussion**

To recap the unconditional view in abstract terms: If we observe a very small *P*-value or posterior probability for a statistical hypothesis **H** about a parameter, the apparent incompatibility could be due to failings of the background assumptions **A** from which that probability was deduced, even if the scientific hypothesis that **H** is supposed to represent is correct (e.g., no treatment effect on the outcome measure). Conversely, if we observe a very large *P*-value or posterior probability for a statistical hypothesis **H** about a parameter, the apparent compatibility could be due to failings of **A**, even if the scientific hypothesis that **H** is supposed to represent is grossly violated. Again, compatibility only measures how consistent **H** is with **A** in light of the data; it says nothing about *why* they appear as consistent or inconsistent as they do. Thus, unconditionally, a very narrow compatibility interval for a targeted parameter means that, only a very narrow range of models that satisfy **A** meets the criterion for “high enough” compatibility for interval inclusion. Interval narrowness may only reflect that **A** represents a visibly unrealistic set of assumptions, rather than that the data provide accurate evidence about the scientific hypothesis corresponding to **H**.

One objection often raised against frequentist statistics is the unreality of the hypothetical repetitions to which error and coverage rates are supposed to apply. But the "long run" which these repetitions are supposed to represent is not necessary under descriptive or information interpretations such as compatibility [3, 17, 36]. Error and coverage rate interpretations convey valid information only when we know the assumption violations would not increase error rates (or reduce coverage rates); otherwise, in the face of assumption uncertainty, these rates become conditional interpretations that are largely irrelevant to sound practice [3, 6, 36, 66].

We thus argue that teaching and practice should emphasize that long-run error and coverage rates are hypothetical ideals which are rarely justified in light of the harsh limitations of social, health, and medical research. While the statistical computations need not change, our recommendation is not simply one of exchanging words, but instead involves a different perspective and description of the computational outputs. With this change, teaching and reporting replaces traditional statistical inferences with unconditional compatibility interpretations. This is done to allow for assumption uncertainty in descriptions of statistical results, rather than only in discussions of those results. This shift in description avoids the overconfident connotations seen in established terms like "significance", "power", "confidence", "support", or "credibility".

To explain our recommendations, we started out by defining the background model logically as the set of all assumptions (constraints) used in deriving a statistic (whether a *P*-value, interval estimate, or posterior probability), including auxiliary assumptions about selection for observation and treatment, distributional forms, etc. This means that all statistical inferences are model dependent, even those from so-called nonparametric or "distribution-free" methods that produce only *P*-values, as well as all Neyman-Pearson ("hypothesis tests" and "confidence intervals"), fiducial, pure-likelihood, and Bayesian methods. They differ greatly however in how explicit they are about their dependence on assumptions.

*Unconditional compatibility is descriptive, not inferential*

Compatibility statements are intended to be more explicit about auxiliary assumptions than are conventional statements. Their advantage is that they can deconditioned (treated as descriptive statements) to avoid making inferences based on uncertain assumptions. In parallel with everyday usage, compatibility (consistency, consonance) and its negation (incompatibility, discrepancy) refer only to some measure of distance or divergence between a fitted model and the numeric data, or between nested models. Unlike Neyman-Pearson decisions and posterior probabilities, the *P*-value is only a statement about the position of a data summary relative to what the model predicts. It thus can be divorced from the "inferential" baggage of decision or betting that attends its usual interpretations. While compatibility statements can also be interpreted in the conventional conditional sense, that interpretation would miss the value added by deconditioning.

To appreciate how much logically weaker compatibility is compared to claims based on "error" rates" and posterior probabilities, consider that everything we have observed is perfectly compatible with the hypothesis (believed by some) that we are actually part of a simulation program in a hypercomputer, with all experiences being delusions generated from feedback between inputs and outputs of our personal subprogram (a modern version of the "brain in a jar" scenario). It is also perfectly compatible with the older theory (believed by many) that the universe was created in six days around seven thousand years ago, and that all its fossils, artefacts, and geologic and cosmologic properties were created then in order to test our faith in the literal interpretation of creation account in the Old Testament, by deceptively pointing to the very different explanation for the empirical world now accepted as “scientific”. There is also a denumerable infinitude of other theories perfectly compatible with all our experiences.

*What about the need for decisions?*

We have been concerned only with how to more validly describe statistical summaries. Decisions based on those summaries are often needed, but statistical decision theory [67] is a massive, deep topic beyond not only our current scope but that of the vast majority of users and consumers of statistics. A key point is that, when background assumptions are uncertain, *neither* conditional nor unconditional interpretations suffice for formal statistical decision methods. Those methods require some type of utility measure or loss function, along with prior distributions (whether empirical or subjective) that incorporate all important uncertainties in the application [68, 69]. Such detailed specifications are often well beyond the temporal and technical resources of researchers.

The conventional dichotomous decision framework (rejecting hypotheses based on whether a *P*-value passes some cutoff, or an interval includes a parameter value) is based entirely on the conditional interpretation, and thus is invalid when that interpretation is questionable, as we argue it usually is in contexts such as health and medical research. It is absurd that the conditional interpretation framework remains promoted as "scientific inference" even in the face of large assumption uncertainties. Use of that framework for critical decisions without attention to actual losses (costs and benefits) from the decisions is at best incompetent and can be potentially catastrophic. The persistence of the conventional framework despite such stark deficiencies reflects the unfamiliarity with and complexity of better methods, and the lack of agreement about simple replacements.

*Going beyond unconditional descriptions to causal bias analysis*

To repeat, unconditional description does not by itself change any model or statistical result, although it may lead to revising the analysis model to a more realistic form. It should be contrasted with sensitivity and bias analyses, in which the analysis model is varied to display sensitivity to modeling choices and to incorporate unmodeled effects on data. Such effects include those of unmeasured factors that influence treatment or exposure and selection into the analysis (both before and after data collection) and effects of misclassification and other mismeasurements. These types of analyses can be intricate and have generated a large literature on these topics (e.g., see citations given in [5–8, 70, 71]).

A major problem for sensitivity and bias analysis is that it multiplies modeling choices ("researcher degrees of freedom"), and itself suffers from extreme sensitivity to those choices. It adds dimensions to the space of models entertained, usually in ways that cannot be checked against the analysis data. This forces the use of prior distributions on the model space to counter this expansion and achieve statistical identification, either in the form of a systematic simple lattice of models (points) in the space (as in standard sensitivity analyses) or in the form of explicit prior distributions on nonidentified parameters (with all the choices and complexities those entail) [5–8, 68-71]. As a result, it is questionable whether such analyses can be reasonably expected be conducted properly by most research teams, and indeed some reviews suggest otherwise [71] .

Sensitivity and bias analyses are thus not an alternative to unconditional interpretations, but rather an attempt to preserve conditional interpretations by combining highly flexible models with more extensive contextual information. In this view, any conventional analysis can be considered one point in a bias analysis, the one in which unmodeled bias sources are assumed to be absent – an assumption rarely, if ever, realistic in observational studies and complex randomized trials in social, health, and medical sciences [70]. Our motivation is thus to provide a simple, noncomputational amendment to conventional analyses without demanding the effort and complexity of bias analysis. We further note that unconditional interpretations can (and we would argue should) be applied to bias analyses as well.

## Conclusions

Under current scientific norms, we don't expect enough critical thinking about statistical summaries and their proper place in the larger process of scientific inference. Those summaries are typically used in conditional interpretations with insufficient sense of how fragile the resulting inferences are in the face of assumption uncertainties. This problem is one reason why many writers have argued that the research literature has given too much authority to those summaries, especially to their use with cutoffs to create arbitrary and often harmful screening devices that are exquisitely sensitive to violations of auxiliary assumptions (our **A**). A major source of the harms is defaulting to traditional conditional interpretations that hide those auxiliary assumptions and their violations from clear view. Deconditioning (giving priority to descriptive rather inferential interpretations of statistics) is one way to allow for the limitations of statistical summaries.

Assumptions should be considered hypothetical statements surrounded by their own uncertainties Unconditional interpretations are a way to immediately acknowledge these uncertainties. While the shift is simple from a technical perspective, it can be initially challenging from a conceptual perspective, partly because of the history and norms around conditional inference embedded into scientific culture and the thinking of practicing scientists today. Unconditional inference is not meant to make the process of scientific inference easier, but with relatively little additional effort it adds honesty and integrity in interpretations – which should be of much higher value than ease of generating "inferences" that hinge on uncertain and often doubtful assumptions.

**Appendix. The scaling and interpretation of binary *S*-values**

A bit is a small unit for measuring information, much like a degree for measuring comfortable room temperature or a centimeter for measuring finger length. A difference of under 1 bit represents an imperceptible difference in information; this means that 4.3 is about the same as 5.1 in practical terms. Anything finer is excessive precision and shown only for computational checking. Thus, for practical purposes, we could round to the nearest integer the binary *S*-value $s = \log_2(1/p) = -\log_2(p)$ (which equals about $-3.32 \cdot \log_{10}(p)$ in base-10 logs or $-1.44 \cdot \ln(p)$ in natural logs). This scaling makes clear that the information difference between $p = 0.07$ and $p = 0.04$ corresponds to $-\log_2(0.07) \approx 4$ minus $-\log_2(0.04) \approx 5$, which is as minor as the difference between $p = 0.87$ and $p = 0.50$, which is $-\log_2(0.87) \approx 0$ minus $-\log_2(0.50) \approx 1$ (a similar point has been made using repeated-sampling "significance" [72], a concept and term we seek to avoid).

We can calibrate our intuitions about a binary *S*-value *s* via thought experiments. For example, suppose we obtain a *P*-value of $p = 0.031$ for the fit of model **M**, which yields an *S*-value of $-\log_2(0.031) \approx 5$. The amount of information against **M** supplied by $p = 0.031$ is then equivalent to the amount of information against the "no-effect" hypothesis supplied by a simple equal-allocation randomized trial with 5 deaths among the treated and 0 deaths among the controls. In more detail, suppose we conduct a simple, perfect randomized trial with equal allocation to treated and control groups, to see the effect of a treatment on a binary outcome (e.g., death vs. survival) that is independent across patients and very uncommon over the study period (a frequency under 1% in each group). Then, upon observing *s* outcomes among the treated vs. none among the controls, the *P*-value for the no-effect hypothesis would be (to close approximation) $\frac{1}{2}^{s}$. Thus, suppose we observe a *P*-value *p* for a model **M** of interest. We may equate the information $s = -\log_2(p)$ that this observation supplies against **M** to the information against a "no-effect" hypothesis that a simple, perfect randomized trial would supply if it observed *s* outcomes and they were all in the treated group. An equivalent calibration can be given in terms of a simple coin-tossing experiment to judge fairness of the tosses [1–3].

Other scales for *S*-values arise from using other bases for the log of *p*; for example, using base-*e* (natural) logs yields units called *nats* and makes the *S*-value a betting score; using base-10 logs makes the *S*-value the *logworth* measure found in machine learning. See [73] for further details.

**Abbreviations: ASD**: Autism spectrum disorder; **CI**: Compatibility interval; **HDPS**: High-dimensional propensity score; **HR**: Hazard ratio; **PH**: Proportional-hazards; ***S***-value: Surprisal (Shannon-information) value

**Declarations:**

**Ethics Approval and Consent to Participate:** Not applicable.

**Consent for Publication:** Not applicable.

**Availability of data and materials:** Not applicable.

**Competing Interests:** The authors declare that they have no competing interests.

**Funding:** This work was produced without funding.

**Authors' Contributions:** All authors have contributed to and approved the submitted manuscript, and have agreed to be personally accountable for their own contributions related to the accuracy of the work.